\documentclass[reprint,  amsmath, amssymb,  aps, prl, nofootinbib]{revtex4-1}
\usepackage{indentfirst}
\usepackage{graphicx}
\usepackage{dcolumn}
\usepackage{bm}
\usepackage{amsmath}
\usepackage{amsfonts}
\usepackage{amssymb}
\usepackage{physics}
\usepackage{braket} 
\usepackage{color}

\begin{document}
\title{Current-driven tricritical point in large-$N_{c}$ gauge theory}
\author{Takuya Imaizumi, Masataka Matsumoto and Shin Nakamura}
\affiliation{Department of Physics, Chuo University, Tokyo 112-8551, Japan}

\begin{abstract}
We discover a new tricritical point realized only in non-equilibrium steady states, using the AdS/CFT correspondence. Our system is a (3+1)-dimensional strongly-coupled large-$N_{c}$ gauge theory. The tricritical point is associated with a chiral symmetry breaking under the presence of an electric current and a magnetic field. The critical exponents agree with those of the Landau theory of equilibrium phase transitions. This suggests that the presence of a Landau-like phenomenological theory behind our non-equilibrium phase transitions.
\end{abstract}

\maketitle

Critical phenomena are important since they show universal behaviors irrespective of the microscopic details of the systems. 
However, critical phenomena in non-equilibrium steady states\,(NESSs) have not yet been investigated comprehensively.
In NESSs, we have an additional control parameter that is absent from equilibrium systems.
For example, the electric current can be a new variable that parameterizes the phase diagrams of the systems of current-driven NESSs.
However, the phase diagrams and the critical phenomena associated with the current have not yet been completely understood. One reason is the lack of theoretical approaches especially for the non-linear regime.

In order to investigate the phase structure of the current-driven NESSs, 
we employ the anti-de Sitter/conformal field theory (AdS/CFT) correspondence. The AdS/CFT correspondence is a duality between a classical gravity theory and a strongly-coupled quantum gauge field theory\,\cite{Maldacena1997,Gubser1998,Witten1998}. 
One of the important applications of this correspondence is to study non-equilibrium physics.\footnote{For example, applications of AdS/CFT correspondence to non-equilibrium physics are reviewed in \cite{Hubeny2010,Kundu2019}} By using the duality, one can analyze the gauge field theory beyond the linear response regime in terms of the gravity theory. For example, the nonlinear conductivity was computed in a NESS in a probe brane model\,\cite{Karch2007}. It has been reported that the probe brane system exhibits interesting behaviors such as the negative differential conductivity\,\cite{Nakamura2010} and the non-equilibrium phase transition\,\cite{Nakamura2012}. It is suggested that critical phenomena of this non-equilibrium phase transition have remarkable similarity to those of the Landau theory of equilibrium phase transitions\,\cite{Matsumoto2018}. A similar non-equilibrium phase transition was also observed in the presence of an external magnetic field\,\cite{AliAkbari2013}.

In this paper, we investigate a spontaneous symmetry breaking of $U(1)$ chiral symmetry of a strongly-coupled large-$N_{c}$ gauge theory under the presence of an electric current density $J$ and a magnetic field $B$ perpendicular to the current. We assume that the system is homogeneous and the size of the system is infinite.
The gauge theory we employ is $SU(N_{c})$ ${\cal N}=4$ supersymmetric Yang-Mills\,(SYM) theory in (3+1) dimensions with ${\cal N}=2$ hypermultiplet. The charged particles belong to the hypermultiplet whereas the particles that form the heat bath belongs to the sector of ${\cal N}=4$ SYM.
The charged particles in our system are massless. This corresponds that we are dealing with gapless systems in condensed matter physics when the system has chiral symmetry.
We find that the system exhibits first-order or second-order phase transitions of  the chiral symmetry breaking. A tricritical point\,(TCP) exists between the line of the first-order phase transition and the second-order critical line. We stress that the newly discovered TCP is located at finite $J$ in the NESS regime on the phase diagram, hence we call it as non-equilibrium tricritical point.
We present numerical results for critical exponents at the critical points. 
Our results imply that the phase transitions of the present system are described by a Landau-like phenomenological model.

We would like to stress that our aim is not to investigate the nature of the above mentioned particular microscopic theory of SYM. The reason why we employ the SYM is to make the duality between the gravity theory and the microscopic theory clear. Our aim is to study the universal properties of non-equilibrium phase transitions that fall into the same category with the system presented here.

{\it Holographic setup.}---The gravity dual of our system is the D3-D7 model\,\cite{Karch2003} with electro-magnetic field\,\cite{Karch2007,Ammon2009}. We briefly review the setup of the gravity dual to make the paper self-contained. The dual geometry is a direct product of a five-dimensional AdS-Schwarzschild black-hole and an $S^{5}$:
\begin{equation}
    ds^{2}= \frac{dz^{2}}{z^{2}}-\frac{1}{z^2}\frac{(1-z^{4}/z_{H}^{4})^{2}}{1+z^{4}/z_{H}^{4}}dt^{2}+\frac{\left( 1+z^{4}/z_{H}^{4} \right)}{z^{2}} d\vec{x}^{2}+ d\Omega_{5}^{2},
\end{equation}
where $z\,\left( 0 \leq z \leq z_{H}\right)$ is the radial direction, $t$ and $\vec{x}=(x^{1}, x^{2}, x^{3})$ are the coordinates for the $3+1$ dimensions of the corresponding gauge theory.
The black-hole horizon is located at $z=z_{H}$ and the boundary is located at $z=0$. The Hawking temperature is given by $T=\sqrt{2}/(\pi z_{H})$, which is identified with the temperature of the heat bath in the gauge theory side. 
The line element of the $S^{5}$ is given by 
    $d\Omega_{5}^{2}=d\theta^{2}+\sin^{2} \theta d\psi^{2} +\cos^{2}\theta d\Omega_{3}^{2},$ where $d\Omega_{3}^{2}$ is the line element of the $S^{3}$ part on which the D7-brane wraps.
The configuration of the D7-brane is given by $\theta(z)\,\left( 0\leq\theta\leq\pi/2 \right)$ and $\psi$. We take $\psi=0$ without loss of generality for our purpose.
For simplicity, we set the AdS radius and the radius of $S^{5}$ to 1. 
The dynamics of the D7-brane is described by the Dirac-Born-Infeld (DBI) action
   $S_{\rm DBI} = -T_{D7} \int d^{8}\xi \mathcal{L}_{D7}$ with 
   $\mathcal{L}_{D7} = \sqrt{-\det \left(  g_{ab}+(2\pi l_{s}^{2}) F_{ab}\right)}$
where $g_{ab}$ is the induced metric 
and $F_{ab}$ is the field strength of the $U(1)$ gauge field on the D7-brane.
$l_{s}$ is the string length and $T_{D7}$ is the tension of the D7-brane given by $T_{D7}^{-1}=(2\pi)^{7}l_{s}^{8}g_{s}$, where $g_{\rm s}$ is the string coupling constant. Since $\theta$ depends only on $z$, the non-trivial element of the induced metric is  $g_{zz}=1/z^{2}+\theta'(z)^{2}$, where the prime denotes the differentiation with respect to $z$.

Let us employ the following ansatz for the gauge fields:\,$A_{1}(t,z)=-Et+a_{1}(z)$, $A_{2}(x^{1})=Bx^{1}$ and $A_{a}$ for $a\neq 1$, $a\neq 2$ are zero. 
Then the non-vanishing components of the field strength $F_{ab}$ are
$F_{01}=-F_{10}=-E$, $F_{12}=-F_{21}=B$, and $F_{1z}=-F_{z1}=a_{1}'(z)$. 
From now, we simply denote $x^{1}$ by $x$, and we set $2\pi l_{s}^{2}=1$ for simplicity.
In our setup, the system is neutral and we have current density only in the $x$ direction.
The expectation value of the current density, which is denoted by $J$, is given by $J=\partial \mathcal{L}_{D7} / \partial  a_{1}'(z)$\,\cite{Karch2007}.
In order to make $J$ to be a control parameter, 
let us perform the Legendre transformation $\tilde{\mathcal{L}}_{D7}=\mathcal{L}_{D7}-J \partial \mathcal{L}_{D7} / \partial  a_{1}'$.
For on-shell, we have
$\tilde{\mathcal{L}}_{D7}=\sqrt{-\frac{g_{zz}}{g_{tt}g_{xx}}} \sqrt{\chi \xi},$
where $\chi = J^{2}+{\cal N}^{2} g_{tt}g_{xx}^{2}\cos^{6}\theta$ and $\xi = g_{tt}g_{xx}^{2}+g_{tt}B^{2}+g_{xx}E^{2}$. Here, $\mathcal{N}=\lambda N_{c} / (2\pi)^{4}$.
In order to guarantee the DBI action to be real, the functions $\chi$ and $\xi$ in the square-root must change the sign at the same location on the $z$ axis, say $z=z_{*}$, simultaneously.\footnote{We can show that $\xi$ has a unique zero at $z=z_{*}$ analytically. We have checked that $\chi$ has a unique zero at $z=z_{*}$ in the parameter range we consider, numerically.
}
Thus, $\xi=0$ and $\chi=0$ are satisfied at $z=z_{*}$, which we call {\it effective horizon}.
The first condition gives us the location of the effective horizon explicitly:
\begin{equation}
z_{*}^{4}=\left( F(e,b)-\sqrt{\left( F(e,b)-1 \right)^{2}-1}-1 \right)z_{H}^{4},
	\label{eq:zstar}
\end{equation}
where
$
F(e,b)=1+e^{2}-b^{2} +\sqrt{\left(e^{2}-b^{2} \right)^{2}+2\left(e^{2}+b^{2}\right)+1}.
$
Here $e$ and $b$ are defined as dimensionless quantities:\,$e=2E/(\pi \sqrt{\lambda}T^{2})$ and $b=2B/(\pi \sqrt{\lambda}T^{2})$, and $\lambda$ is the 't Hooft coupling of the gauge theory.
The second condition provides the explicit form of the current density
	$J^{2}=-{\cal N}^{2}g_{tt}g_{xx}^{2}\cos^{6}\theta(z_{*}).$
The D7-brane embedding function $\theta(z)$ in the vicinity of the boundary can be expanded as
    $\theta(z)=\theta_{0} z + \theta_{2} z^{3} + \cdots$.
Here, $\theta_{0}$ and $\theta_{2}$ are, respectively, related to the mass of the charge carriers $m$ and the chiral condensate $\braket{\bar{q}q}$ as 
$m=\theta_{0}$ 
and $\braket{\bar{q}q} =-\mathcal{N} \left( 2\theta_{2}-m^{3}/3\right)$.

{\it Non-equilibrium phase transition.}--- 
We set $z_{H}=1$ for simplicity\footnote{In our convention, the unit of $B$ and $E$ is $\left[(\sqrt{2}T/\pi)^{2} \right]$ and that of $J$ and $\expval{\bar{q}q}$ is $\left[{\cal N}(\sqrt{2}T/\pi)^{3} \right]$ in the following figures.}.
In order to solve the equation of motion\,(EOM) for $\theta(z)$, we have to specify the boundary condition at the effective horizon. 
The explicit form of the boundary condition is given by $\theta'(z_{*})=\left( C-\sqrt{C^{2}+D^{2}} \right) / (z_{*}D)$, where $C=3+2z_{*}^{4}+3z_{*}^{8}$ and $D=3(z_{*}^{8}-1)\tan{\theta(z_{*})}$\,\cite{Nakamura2010,Albash2007}.
$\theta(z_{*})$ is a parameter to be set and $z_{*}$ is determined by Eq.\,(\ref{eq:zstar}) for given $E$ and $B$. We solve the EOM numerically to obtain the solutions $\theta(z)$ in $\varepsilon \leq z \leq z_{*}-\varepsilon_{\mbox{\tiny \it IR}}$, where $\varepsilon$ and  $\varepsilon_{\mbox{\tiny \it IR}} $ are cutoffs introduced to avoid numerical divergence at the boundary and the effective horizon, respectively.

\begin{figure}[tbp]
\centering \includegraphics[width=8.6cm, bb=0 70 958 450]{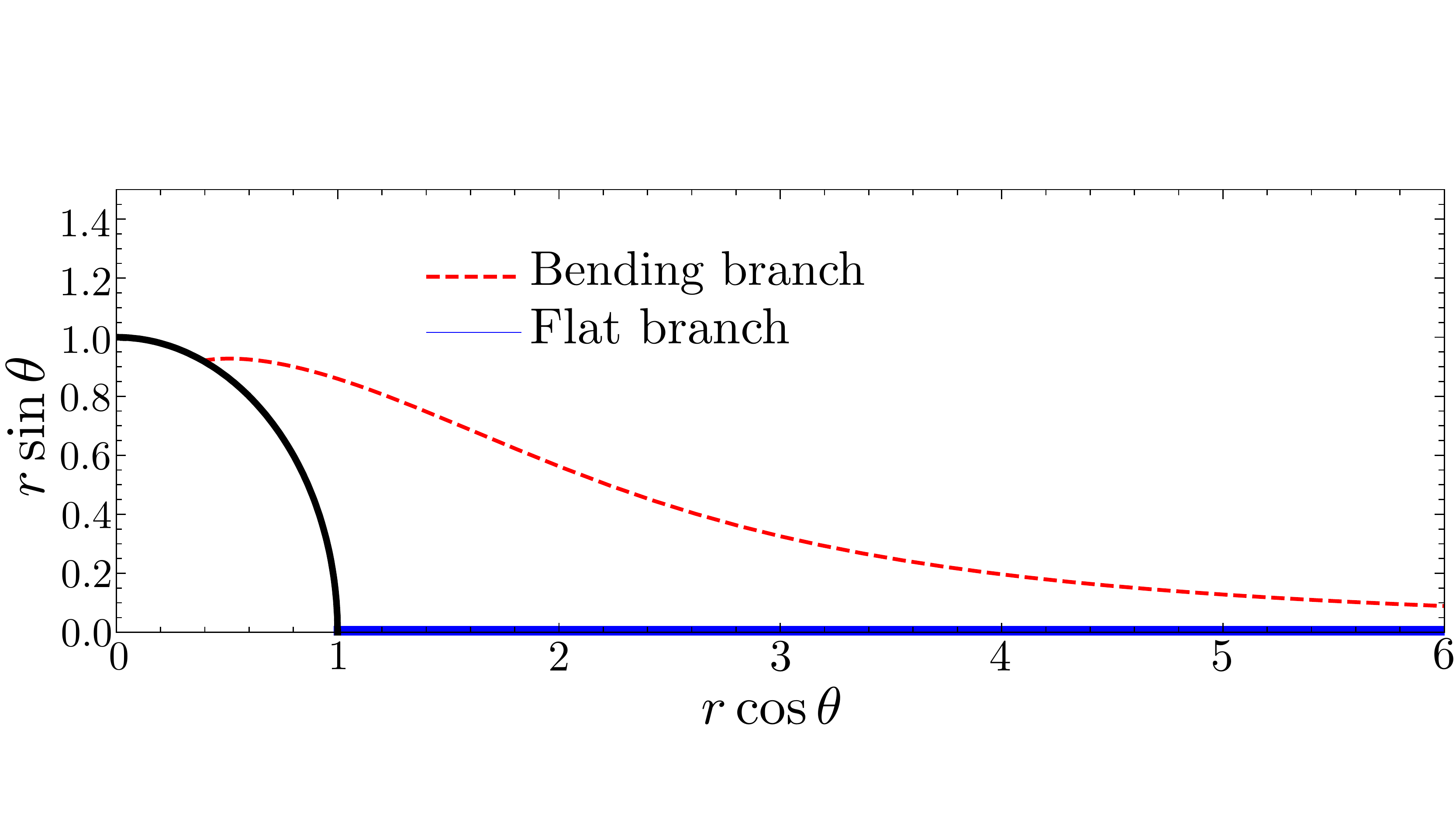}
\caption{The D7-brane configurations of two possible branches for $E=1$ and $B=14$. The black solid line represents the black-hole horizon. The coordinate is changed following by $1/z^{2}=r^{2}+\sqrt{r^{4}-1}$. The black-hole horizon is located at $r=1$, the effective horizon is located at $r=r_{*}>1$, and the AdS boundary is located at $r=\infty$.}
\label{fig:config}
\end{figure}

Once $\theta(z)$ is obtained for each parameter, one can compute $m$ and $\expval{\bar{q}q}$. Inversely, the configurations of the D7-brane and the parameters can be found for the fixed $m$. In this study, since we are interested in the system of massless charge carriers and the spontaneous chiral symmetry breaking, we take solutions for $m=0$ by using the shooting method. We show the configurations of the D7-brane in Fig.\,\ref{fig:config}. There exist two possible branches:\,the flat branch and the bending branch as shown in Fig.\,\ref{fig:config}. The flat branch is a trivial solution:\,$\theta(z)=0$. The bending branch is a non-trivial solution:\,$\theta(z)\neq 0$. These branches can be distinguished by the chiral condensate:\,$\expval{\bar{q}q}=0$ in the flat branch, and $\expval{\bar{q}q}\neq 0$ in the bending branch. In other words, the chiral symmetry is preserved in the flat branch, whereas it is spontaneously broken in the bending branch. 
To be more specific, the chiral symmetry is the $U(1)_{R}$ symmetry of the SYM theory\,\cite{Babington2004}. 
Compared to the previous studies on the chiral symmetry breaking in the AdS/CFT correspondence \cite{Babington2004,Filev2007,Albash2008,Filev2008,Erdmenger2007}, the crucial difference in our study is that we consider chiral symmetry breaking in the NESS regime where $J\cdot E\neq 0$.

\begin{figure}[tbp]
\centering
\includegraphics[width=8.6cm, bb=0 50 960 540]{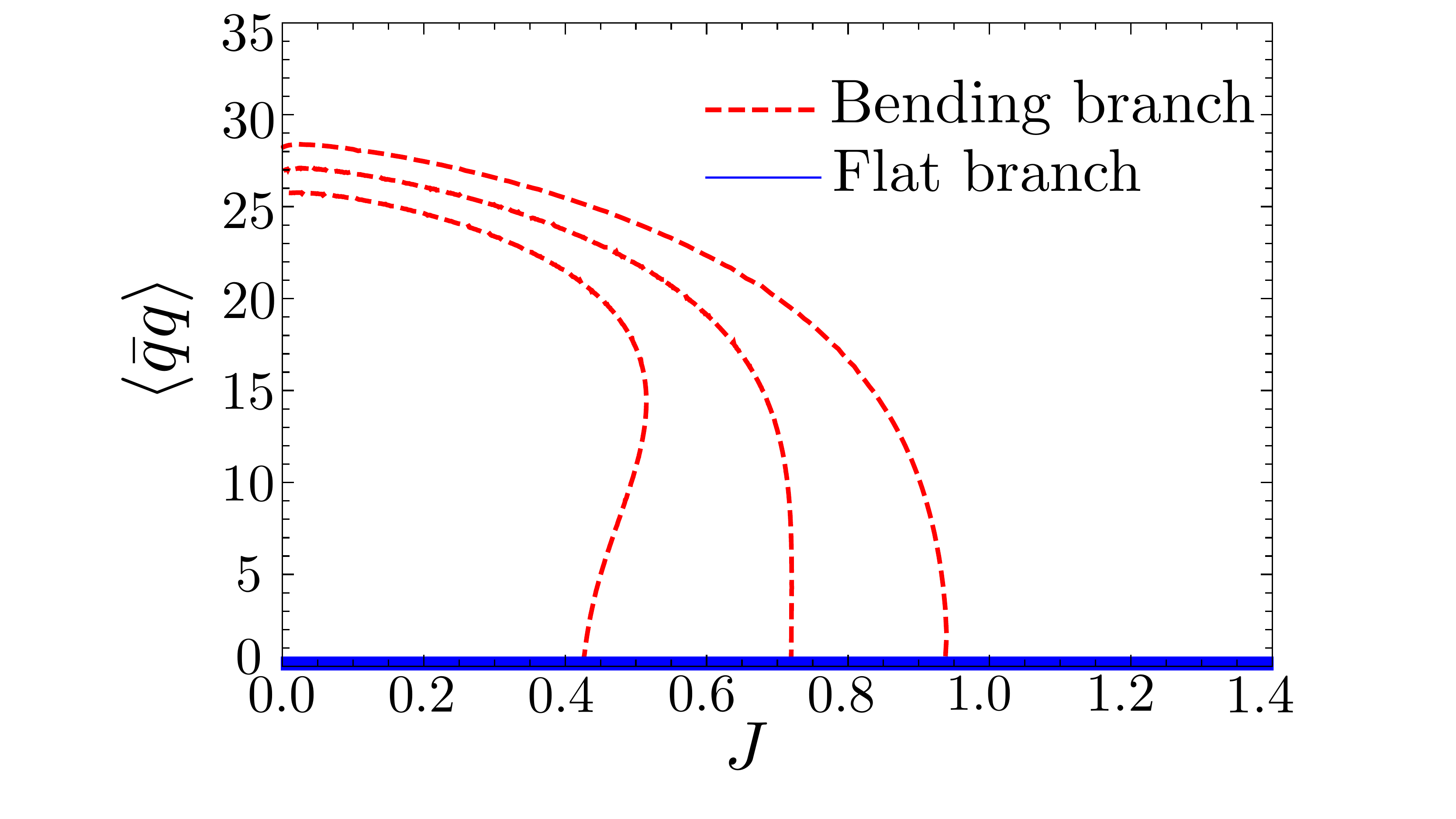}
\caption{The behaviors of the chiral condensate $\braket{\bar{q}q}$ as a function of the current density $J$ for $B=19$, $B=20$, and $B=21$ from left to right.}
\label{fig:Jqq}
\end{figure}

Fig.\,\ref{fig:Jqq} shows the behaviors of the chiral condensate $\expval{\bar{q}q}$ and the current density $J$ for various $B$. Note that if we choose the current density as a control parameter, there is a multivalued region where $\expval{\bar{q}q}$ has two or three possible values for a given $J$. In this region, we expect that only the most stable solution is realized.

In the following, we assume that the most stable solution has the lowest {\it thermodynamic potential} which is defined by the renormalized Hamiltonian of the D7-brane per unit volume with appropriate Legendre transformation \cite{Nakamura2012,Matsumoto2018}\footnote{See also \cite{Banerjee:2015cvy}, for a similar proposal.}$^{,}$\footnote{The Hamiltonian at $J\rightarrow 0$ gives the free energy in equilibrium. The thermodynamic potential employed in Ref.\,\cite{Evans2010,Evans2011m,Evans2011} can be understood as the Legendre transform of the Hamiltonian to make the electric field a control parameter.}.
The Hamiltonian density is given by
\begin{eqnarray}
\tilde{\mathcal{H}}_{D7}=g_{xx}\sqrt{-g_{tt}g_{zz}}\sqrt{\frac{J^{2}+\mathcal{N}^{2} g_{tt}g_{xx}^{2} \cos^{6} \theta}{g_{tt}g_{xx}^{2}+g_{tt}B^{2}+g_{xx}E^{2}}}.
\end{eqnarray}
Then the thermodynamic potential (per unit volume) is defined as 
\begin{equation}
    \tilde{F}_{D7}(B, J) =\lim_{\varepsilon\rightarrow 0} \left[ \int_{\varepsilon}^{z_{*}} dz \tilde{\mathcal{H}}_{D7} -L_{\mbox{\tiny count}}(\varepsilon)\right],
    \label{eq:TP}
\end{equation}
where $L_{\mbox{\tiny count}}(\varepsilon)$ is the counterterm to
renormalize the divergence at the boundary $z=0$. The explicit form is given in \cite{Karch2007} except for the finite counterterm we propose\footnote{See also \cite{Karch2006}, for the holographic renormalization of probe D-branes.}. The details on the finite counterterm is shown in the supplemental material.

If we compute the thermodynamic potential in each plot of Fig.\,\ref{fig:Jqq}, we find that as the magnetic field increases, the thermodynamic potential of the bending branch becomes smaller and it eventually becomes more stable than the flat branch. As a result, there is a transition between these two branches.
We regard this transition as a {\it non-equilibrium phase transition}.
More interestingly, when $B$ becomes larger, the chiral condensate of the bending branch changes from a multi-valued function to a single-valued function, and thus the discontinuous jump of the chiral condensate becomes continuous transition. Hence, the {\it first-order phase transition} changes to the {\it second-order phase transition} by increasing $B$.
In this paper, we choose the chiral condensate $\braket{\bar{q}q}$ as an order parameter.

\begin{figure}[tbp]
\includegraphics[width=8.6cm, bb=0 10 960 540]{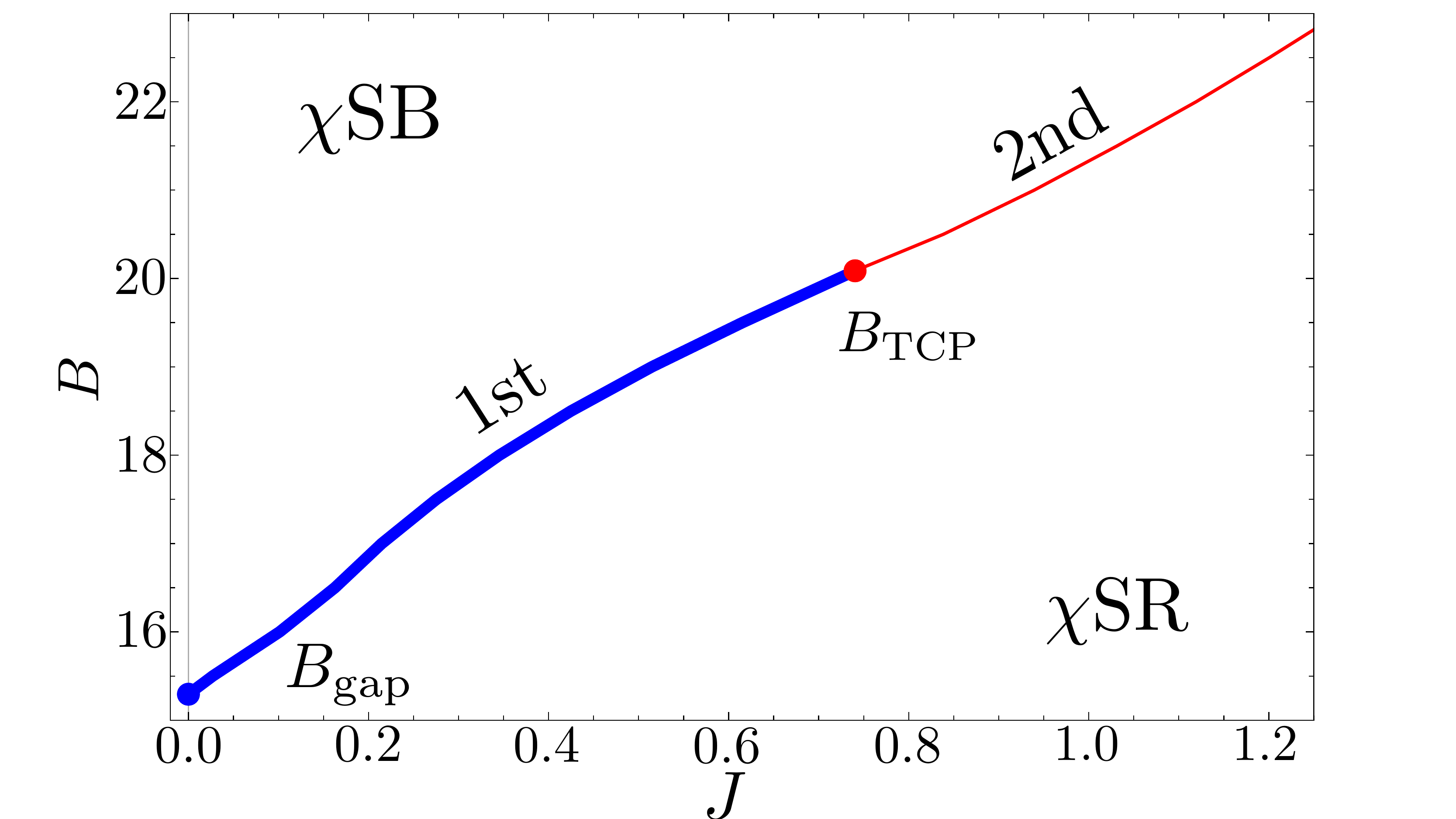}
\caption{Phase diagram of the present system on the slice of $m=0$. There are two specific points denoted by $B_{\rm gap}$ and $B_{\rm TCP}$. The first-order phase transitions occur along the thick blue curve and the second-order phase transitions occur along the red curve. These curves completely separate the phase with chiral symmetry restoration ($\chi{\rm SR}$) and that with chiral symmetry breaking ($\chi{\rm SB}$).}
\label{fig:Phase}
\end{figure}

\begin{figure}[tbp]
\includegraphics[width=8.6cm,bb=50 50 900 540]{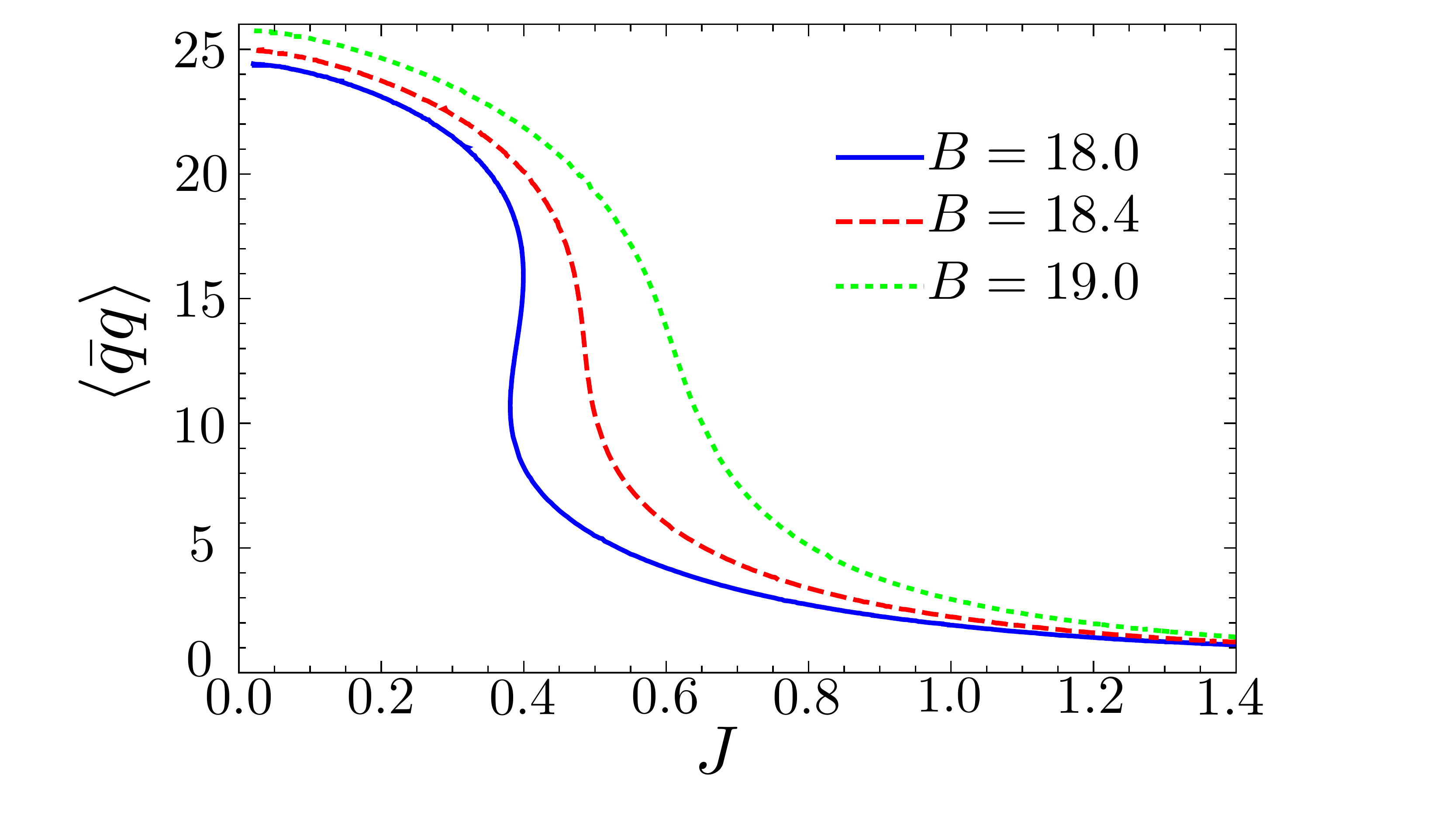}
\caption{The behavior of the chiral condensate $\expval{\bar{q}q}$ at $m=0.01$. When the magnetic field increases, the transition of the chiral condensate appears to be changed from first-order\,(solid line) to second-order\,(dashed line) and crossover\,(dotted line).}
\label{fig:JqqM}
\end{figure}

{\it Phase diagram.}---
To summarize the foregoing discussions, we show the phase diagram of the non-equilibrium phase transitions in Fig.\,\ref{fig:Phase}. The thick blue curve is the first-order phase transition line and the red curve is the critical line where the second-order phase transition occurs. Note that there are two specific points in the phase diagram:\,$B_{\rm gap}$ and $B_{\rm TCP}$. For $B < B_{\rm gap}$, stable solutions are only the flat branches. However, for $B > B_{\rm gap}$, the bending branch can be stable.
The other specific point $B_{\rm TCP}$ is the {\it tricritical point}\,(TCP) that is defined as a point where three coexisting phases become identical simultaneously. In order to see this in our system, it is necessary to consider another parameter dimension:\,the mass of the charged particles $m$ which breaks the chiral symmetry explicitly.
In analogy with the QCD phase diagram\footnote{For example, see \cite{Halasz1998}.},
we find that the $(B, J)$ phase diagram with a finite mass contains a first-order phase transition line which terminates at a critical point. This can be understood by checking the behaviors of the chiral condensate as a function of the current density for small mass as shown in Fig.\,\ref{fig:JqqM}. Critical points for different masses form a critical line in the $(B, J, m)$ phase diagram and the critical line terminates at $B_{\rm TCP}$ that is located on the slice of $m=0$.
We can extend the phase diagram into the negative mass region by performing a $U(1)$ chiral transformation. One finds that the extended phase diagram is symmetric under 
$m \rightarrow -m$. Hence, the three critical lines end at a single point $B_{\rm TCP}$. This is why $B_{\rm TCP}$ is regarded as the tricritical point.

\begin{figure}[tbp]
\centering
\includegraphics[width=8cm, bb=0 0 910 540]{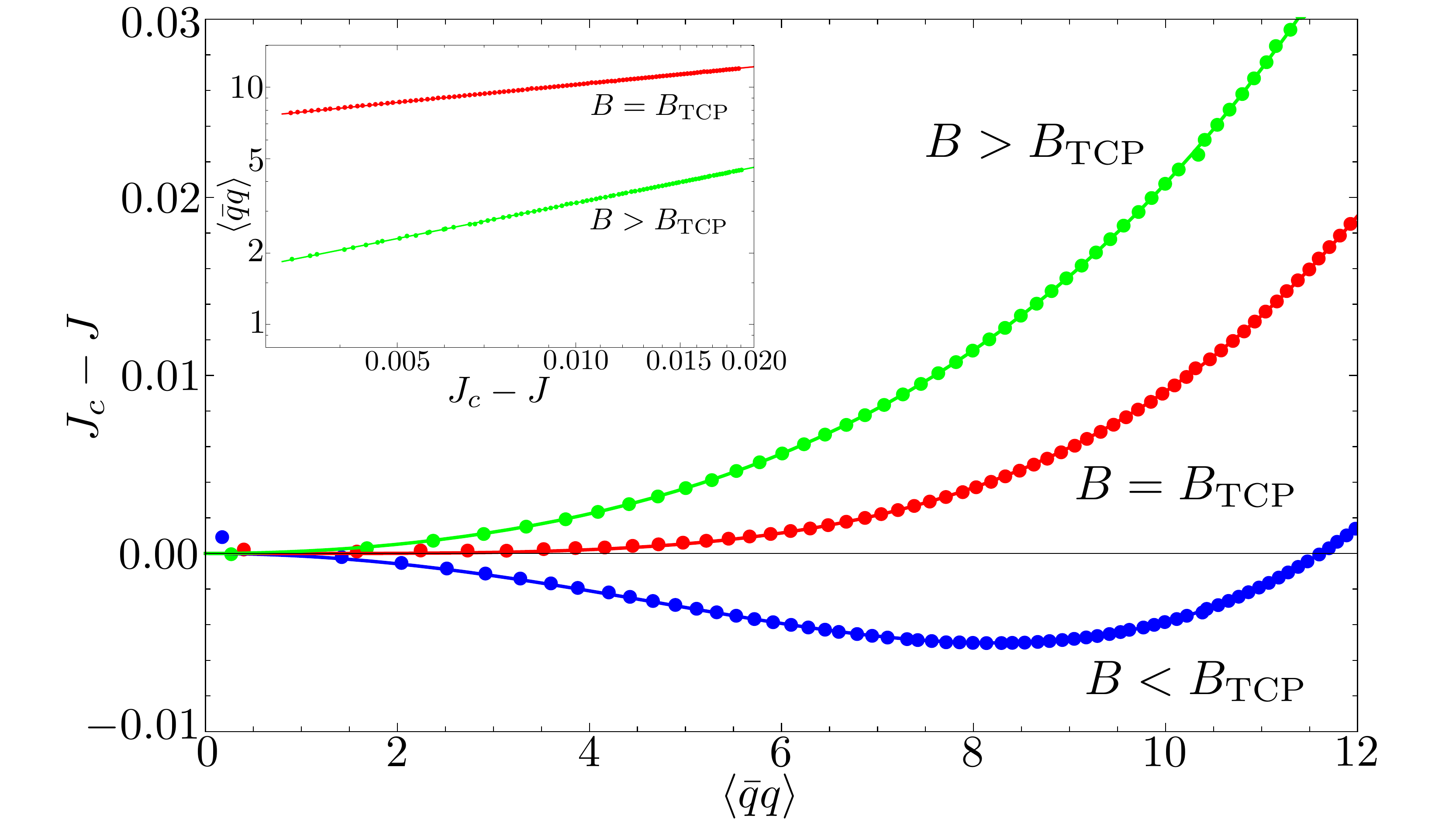}
\caption{The relationship between the chiral condensate $\expval{\bar{q}q}$ as a function of the current density $J_{c}-J$. The numerical plots and the lines obtained by non-linear fitting to the form of (\ref{eq:4th}) for $B=19.8\,$(Blue), $B=B_{\rm TCP}\,$(Red), and $B=20.4\,$(Green) are shown. The inset shows numerical plots and linear fittings for $B=B_{\rm TCP}$ and $B=30$.}
\label{fig:qqvsJ}
\end{figure}

{\it Critical Exponent.}---
In the previous section, we found second-order phase transitions in our non-equilibrium system. In order to understand 
the critical phenomena, we analyze the critical exponents. Critical exponents are defined from the behavior of the order parameter near the critical point. In our analysis, we define the critical exponent $\beta$ by \footnote{In Ref.\,\cite{Matsumoto2018}, the critical exponent related to the chiral condensate and the current density is defined as $\abs{\expval{\bar{q}q}-\expval{\bar{q}q}_c} \propto \abs{J-J_{c}}^{1/\delta}$. In this study, we use $\beta$ since the non-equilibrium phase transition in question is different.} $\braket{\bar{q}q} \propto \left( J_{c} -J \right)^{\beta}$ along the line of $m=0$\footnote{The line $m=0$ ($J \leq J_{c}$) is the line of the first-order phase transition on the phase diagram of $J$-$m$ plane. Note that this is solely determined by the symmetry of the system.}, where $J_{c}$ is the current density at the critical point.  
Fig.\,\ref{fig:qqvsJ} shows the behaviors of $\braket{\bar{q}q}$ for $B>B_{\rm TCP}$, $B=B_{\rm TCP}$, and $B<B_{c}$. As shown in the inset of Fig.\,\ref{fig:qqvsJ}, we find $\beta=1/2$ for $B>B_{\rm TCP}$ and $\beta=1/4$ for $B=B_{\rm TCP}$ within the numerical error by linear fitting. 

\begin{figure}[tbp]
\centering
\includegraphics[width=6cm, bb=0 0 334 691]{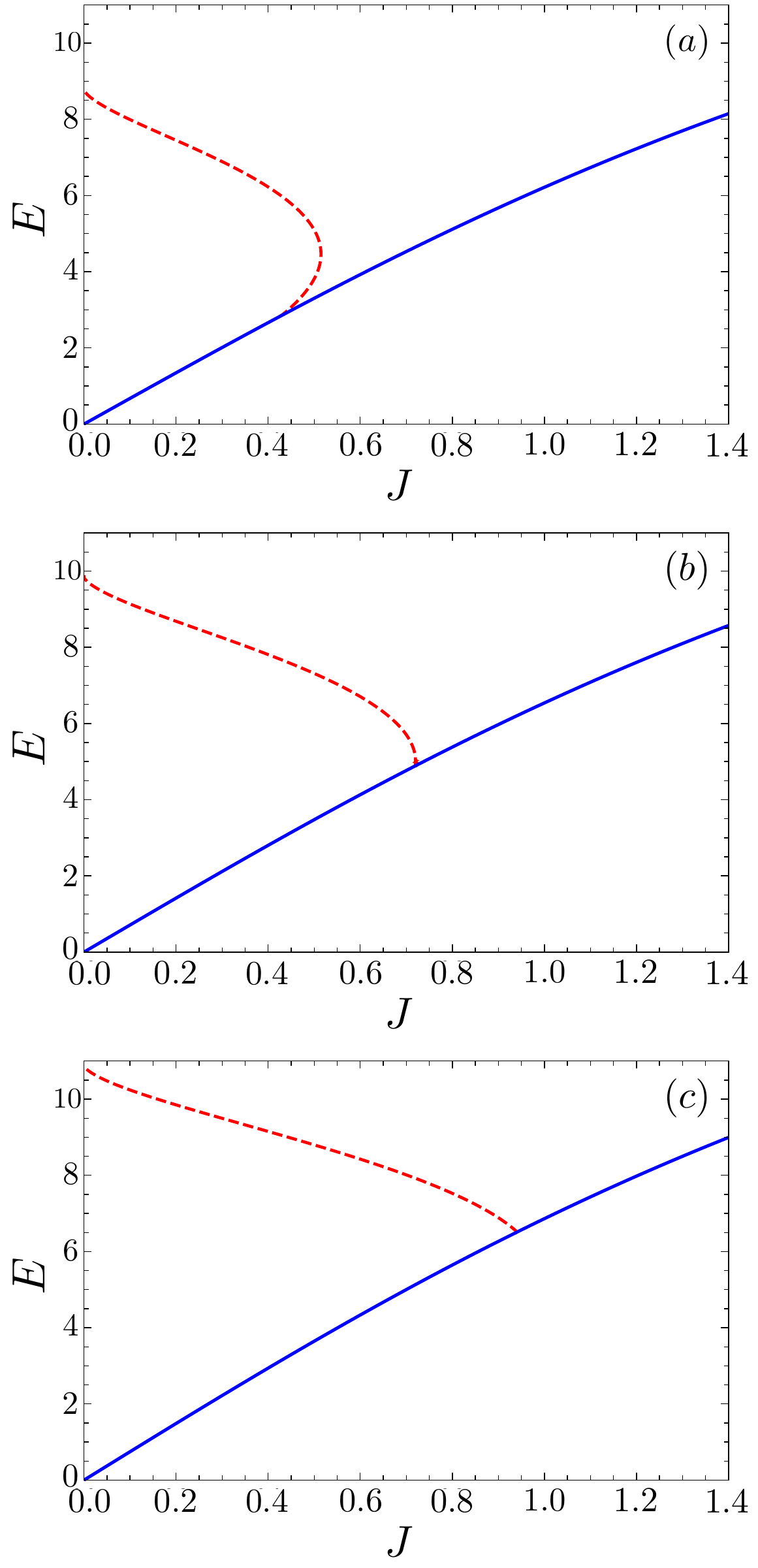}
\caption{The $J$-$E$ characteristics for $(a)\,B=19$, $(b)\,B=20$, and $(c)\,B=21$.}
\label{fig:JE}
\end{figure}

{\it Analytical approach.}---
We show that the values of the critical exponents can be derived analytically as follows.
Let us focus on the case for $B \geq B_{\rm TCP}$. 
In the language of the D7-brane configuration, $\theta(z)$ continuously changes from $\theta(z)=0$ to $\theta(z)\neq 0$ when the chiral symmetry is broken in the second-order phase transition. This means that 
$\theta(z_{*}) \ll 1$ in the vicinity of the critical point. 
Since $J = \left. \sqrt{-g_{tt}g_{xx}^2} \right|_{z_{*}} \cos^3{\theta(z_{*})}$ is an even function of $\theta(z_{*})$, $J$ can be expanded as 
\begin{eqnarray}
J=J_{c}+ a_{2} \theta(z_{*})^{2} + a_{4} \theta(z_{*})^{4} + \cdots,
\label{J-expand}
\end{eqnarray}
where $a_{i}(i=2,4,\cdots)$ are coefficients that depend on $z_{*}$. Here we have used the fact that $J=J_{c}$ at the critical point where $\theta(z_{*})=0$. Note that $J_{c}$ depend only on $B$.

Next, we argue that the chiral condensate is an odd function of $\theta(z_{*})$. When we flip the sign of $\theta(z_{*})$, the function $\theta(z)$ also flips the overall sign. For massless case, $\theta(z)$ in the vicinity of the boundary is expanded as $\theta(z)=-\frac{1}{2}\braket{\bar{q}q}z^{3}+O(z^{5})$. Then we find that the chiral condensate flips the sign, too. Therefore, the chiral condensate is an odd function of $\theta(z_{*})$ which may be expanded as 
$\braket{\bar{q}q}=b_{1} \theta(z_{*}) +b_{3} \theta(z_{*})^{3} +\cdots$,
where $b_{i}(i=1,\,3,\,\cdots)$ are coefficients that depend on $z_{*}$. 

Suppose that $b_{1}\neq 0$. Then the chiral condensate is {\em linear} in $\theta(z_{*})$ in the vicinity of the critical point, and (\ref{J-expand}) gives
\begin{equation}
    J_{c}-J = c_{2} \braket{\bar{q} q} ^{2} + c_{4} \braket{\bar{q} q} ^{4} + \cdots,
    \label{eq:4th}
\end{equation}
where $c_{2}$ and $c_{4}$ are some coefficients.  The numerical results given in Fig.\,\ref{fig:qqvsJ} suggest that $c_{2}=\kappa\left( B-B_{\rm TCP} \right)$ with $\kappa>0$ and $c_{4}>0$, in the vicinity of $B=B_{\rm TCP}$. We have checked numerically that $c_{2}$ is indeed proportional to $B-B_{\rm TCP}$ and $c_{4}$ is a small positive value in the vicinity of the critical point.
Then we obtain $\braket{\bar{q} q} \sim (J_{\small c}-J)^{1/2}$ for $B>B_{\rm TCP}$ and $\braket{\bar{q} q} \sim (J_{\small c}-J)^{1/4}$ for $B=B_{\rm TCP}$, respectively. The above mentioned numerical results for the critical exponents justify the above consideration. Therefore, we conclude that $\beta=1/2$ for $B>B_{\rm TCP}$ and $\beta=1/4$ for $B=B_{\rm TCP}$.

{\it Conductivity.}---
In \cite{Nakamura2012, Matsumoto2018}, it has been found that the conductivity $\sigma=J/E$ also plays a role of order parameter at least for the purpose of detection of the critical phenomena. From experimental point of view, measurement of conductivity is much more easier than that of chiral condensate. Therefore, it is worth while studying the phase transitions from the viewpoint of conductivity. 
Fig.\,\ref{fig:JE} shows the $J$-$E$ characteristics for various $B$. 
In terms of the conductivity, our non-equilibrium phase transitions are regarded as the transition between the negative differential conductivity\,(NDC) phase and the positive differential conductivity\,(PDC) phase, which are corresponding to two branches, the bending branch and the flat branch, respectively. 
As shown in Fig.\,\ref{fig:JE}, the transition between the PDC phase and the NDC phase is also the second-order phase transition along the critical line and the first-order phase transition below $B_{\rm TCP}$. At the tricritical point, $\partial E/\partial J \rightarrow \infty$, which is a similar behavior to that of the chiral condensate in Fig.\,\ref{fig:Jqq}. This shows that the chiral symmetry breaking is closely related to the transition between the PDC phase and the NDC phase.
In other words, our results suggest that our non-equilibrium phase transitions and the tricritical point will be experimentally observed as transitions between the NDC phase and the PDC phase in a strongly coupled system with a constant current.

{\it Conclusion.}---
We studied the phase structure associated with the chiral symmetry of the current-driven NESS by using the AdS/CFT correspondence. We have discovered the current-driven tricritical point.
If we define the critical exponent $\beta$ as $\braket{\bar{q}q} \propto \left( J_{c} -J \right)^{\beta}$, we obtain $\beta=1/2$ on the critical line and $\beta=1/4$ at the tricritical point. These values of the critical exponents agree with those of the Landau theory for equilibrium systems.
We infer that this agreement is due to the large-$N_{c}$ limit in the AdS/CFT correspondence.
Our results suggest the existence of an effective theory for our non-equilibrium phase transitions whose ``Landau free energy'' $f(\langle \bar{q}q\rangle)$ is given by
$
f(\langle \bar{q}q\rangle)
=a(J-J_{c})\langle \bar{q}q\rangle^{2}+b(B-B_{\rm TCP})\langle \bar{q}q\rangle^{4}+c\langle \bar{q}q\rangle^{6}-d m\langle \bar{q}q\rangle,
$
where $a$, $b$, $c$ are positive constants and $d$ is a non-zero constant at a fixed $T$.
It is interesting to investigate the validity of the foregoing ``Landau free energy'' and if other critical exponents also agree with those of the Landau theory or not. These are left for future exploration. 
We propose that this type of non-equilibrium phase transitions and the current-driven tricritical point may be detected in a system consisted of gapless chiral fermions, such as a Weyl semimetal, with a constant current.
We hope the results presented here provide clues to reveal universal properties of some class of NESSs.

\begin{acknowledgements}
The authors are grateful to Y. Fukazawa, H. Hoshino, S. Ishigaki, S. Kinoshita, H. Takakuma and R. Yoshii for helpful discussions and comments. The work of M.M. is supported by the Research Assistant Fellowship of Chuo University. The work of S.N. is supported in part by JSPS KAKENHI Grant Numbers JP16H00810, JP19K03659, JP19H05821, and the Chuo University Personal Research Grant. 
The authors thank the Yukawa Institute for Theoretical Physics at Kyoto University. Discussions during the YITP workshop YITP-W-19-09 on ``Thermal Quantum Field Theory and Their Applications'' were useful to complete this work.
\end{acknowledgements}

\clearpage

\section{Supplemental Material for \\
``Current-driven tricritical point in large-$N_{c}$ gauge theory"}

\setcounter{figure}{6}
\begin{figure*}[b]
\centering
\includegraphics[width=17.2cm, bb=0 80 540 700]{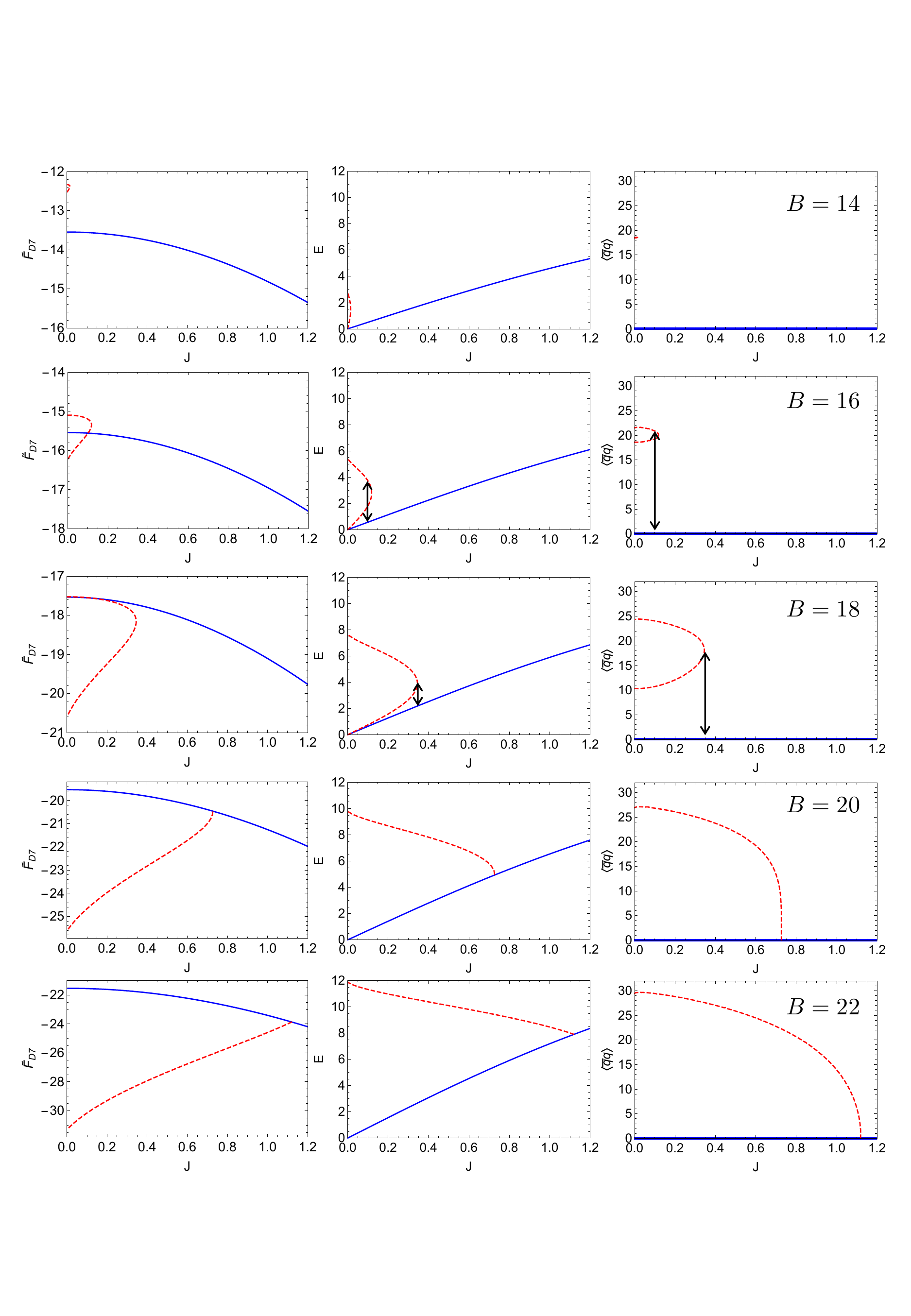}
\caption{Figures in the left column show the thermodynamic potential $\tilde{F}_{\small D7}$ as a function of $J$ for $B=14$, $B=16$, $B=18$, $B=20$, and $B=22$ from top to bottom. In the middle column, $J$-$E$ characteristics are shown for each value of $B$. In the right column, we show $\expval{\bar{q}q}$ as a function of $J$ for each value of $B$. The red dashed plots denote the bending branch and the solid blue plots denote the flat branch. The arrows in the plots indicate the points where the first-order phase transition occurs.}
\label{fig:JEH}
\end{figure*}

In the supplemental material, we show the explicit form of the counterterm in Eq.\,(4) in the main text and the behaviors of the thermodynamic potential. 

$L_{\mbox{\tiny count}}(\varepsilon)$ in Eq.\,(4) is the counterterm that renormalize the divergence at the boundary $z=0$. $L_{\mbox{\tiny count}}(\varepsilon)$ consists of four components
\begin{equation}
    L_{\mbox{\tiny count}}(\varepsilon)=L_{1}+L_{2}+L_{f}+H_{F},
    \label{eq:CT}
\end{equation}
where the first three terms are given in \cite{Karch2007}:
 \begin{eqnarray}
     L_{1} &=&\frac{1}{4}\sqrt{-\det \gamma_{ij}}=\frac{1}{4}\varepsilon^{-4}, \\
     L_{2} &=&-\frac{1}{2}\sqrt{-\det \gamma_{ij}}\theta(\varepsilon)^{2}=-\frac{1}{2}\varepsilon^{-2}\theta_{0}^{2}, \\
     L_{f} &=&\frac{5}{12}\sqrt{-\det \gamma_{ij}}\theta(\varepsilon)^{4}=\frac{5}{12}\theta_{0}^{4}.
     \label{eq:LF}
 \end{eqnarray}
Here $\gamma_{ij}$ is the induced metric on the $z=\varepsilon$ slice.
Our aim is to determine the explicit form of $H_{F}$ for the massless systems of the D3-D7 model. As discussed in\,\cite{Nakamura2012}, $H_{F}$ is essentially given by the Legendre transform of the $L_{F}$ presented in \cite{Karch2007}. However, the $L_{F}$ in \cite{Karch2007} in the present notation is $L_{F}=-\frac{1}{2}F^{2}\log \varepsilon$, and the argument of the logarithm is not explicitly dimensionless. Our claim is that the argument of the logarithm should be made dimensionless by using $F^{2}$ for the purpose of the present work.

In order to specify the explicit form of $L_{F}$ for us, we consider the simplest case. Since renormalization of UV divergence is not affected by the macroscopic setup, we consider the case that the current density is absent and the system is in equilibrium. We achieve this by introducing only a magnetic field, but we write $F^{2}=2(B^{2}-E^{2})$ to make the expressions Lorentz invariant. Then the Hamiltonian gives the equilibrium free energy. Note that this is equal to $(-1)$ times the Lagrangian when the system is static and $E$ is absent. We take the zero-temperature limit,
and we set $m=0$.

Let us evaluate the Lagrangian explicitly.
We introduce coordinates $(\rho, y)$ given by $1/z^{2}=\rho^{2}+y^{2}$ \,{\it i.e.},  
$\rho=\cos\theta(z)/z$ and $y=\sin\theta(z)/z$.
If we expand $\theta(z)$ with respect to $z$, we find that the value of $y$ at the boundary gives the mass of the charged particles:\,$y\to m$ at $\rho \to \infty$. Notice that $\rho$ has a mass dimension in our convention.
%
 
Then the Lagrangian of D7-brane (per unit volume) is given by
\begin{equation}
    L_{D7}=\int^{\infty}_{0}d\rho \rho^{3}\sqrt{1+F^{2}/\rho^{4}}.
\end{equation}
If we introduce cutoffs $\rho_{\rm max}$ and $\rho_{\rm min}$ at the boundary and the origin, respectively, the Lagrangian density is rewritten as
\begin{eqnarray}
    L_{D7}&=&\int^{\rho_{\rm max}}_{\rho_{\rm min}}d\rho\: \rho^{3} \sqrt{1+F^{2}/\rho^{4}} \nonumber \\
    &=&\left[\frac{1}{4}\rho^{2}\sqrt{F^{2}+\rho^{4}}+\frac{1}{4}F^{2} \log \left( \rho^{2}+\sqrt{F^{2}+\rho^{4}} \right)\right]_{\rho_{\rm min}}^{\rho_{\rm max}} \nonumber\\
    &\simeq & \frac{1}{4}\rho_{\rm max}^{2}\sqrt{F^{2}+\rho_{\rm max}^{4}}+\frac{1}{4}F^{2}\log\frac{ \rho_{\rm max}^{2}+\sqrt{F^{2}+\rho_{\rm max}^{4}} } {\rho_{\rm min}^{2}+\sqrt{F^{2}+\rho_{\rm min}^{4}}} \nonumber\\
    &\simeq& \frac{\rho_{\rm max}^{4}}{4}
+\frac{F^{2}}{8}\log\left(\frac{4e}{F^{2}}\rho_{\rm max}^{4} \right),    
\label{LD7-2}
\end{eqnarray}
where $e$ is the Napier number.
Notice that $\rho_{\rm max}$ is related to the UV cutoff $z=\varepsilon$ in the $z$ coordinate as $\rho_{\rm max}=1/\varepsilon$. Here we approximate $\sqrt{F^{2}+\rho_{\rm max}^{4}} \simeq \rho_{\rm max}^{2}\left( 1+\frac{F^{2}}{2\rho_{\rm max}^{4}} \right)$ and ignore the terms that go to zero in the limit $\rho_{\rm max}\rightarrow\infty$ and $\rho_{\rm min} \rightarrow 0$. We note that for finite $F^{2}$, $\rho_{\rm min}$ is harmlessly taken to be $0$. 
The first term is renormalized by $L_{1}$, and the second term may be renormalized by $L_{F}=\frac{F^{2}}{2}\log \rho_{\rm max}$ given in \cite{Karch2007}. 
However, in this case, there is a remaining finite contribution from the second term which causes divergence in the susceptibility
$\left. \partial^{2} H / \partial B^{2} \right|_{B=0,E=0}=-\left. \partial^{2} L_{D7} / \partial B^{2} \right|_{B=0,E=0}$ at the zero-field limit. In order to avoid the divergence in the susceptibility, we need to subtract the finite contribution together with the logarithmic UV divergence. 
We have an ambiguity to choose the finite contribution which does not cause the divergence in the susceptibility. We set this finite term so that the susceptibilities are zero for the vacuum state where $T=F^{2}=0$. Then one finds that the second term in (\ref{LD7-2}) has to be subtracted entirely.
We do not put any further finite term which is independent of $F^{2}$, because we need to make the Hamiltonian zero for the supersymmetric setup of $T=F^{2}=0$. Then we conclude that our $L_{F}$ should be 
\begin{equation}
L_{F}=\frac{F^{2}}{8}\log (4e\rho_{\rm max}^{4}/F^{2})
=-\frac{F^{2}}{8}\log (F^{2}\varepsilon^{4}/4e).
\end{equation}
Now, the argument in the logarithm is made dimensionless.
So far, we have switched off the electric field. However, $E^{2}$ can be safely revived in $F^{2}$ since the renormalization must to be made in a Lorentz invariant way.

The counterterm for the Hamiltonian is straightforwardly obtained by performing Legendre transformation $H_{F}=-L_{F}+E\frac{\partial L_{F}}{\partial E}$. The result is
\begin{eqnarray}
H_{F}=\frac{F_{\rm E}^{2}}{8}\log (F^{2}\varepsilon^{4}/4e)+\frac{E^{2}}{2},
\end{eqnarray}
where $F_{\rm E}^{2}=2(E^{2}+B^{2})$.
One finds that the last term comes from the Legendre transformation.

We compute the renormalized thermodynamic potential using the counterterms given above.
The behaviors of the thermodynamic potential as a function of $J$ for several values of $B$ are shown in the left column of Fig.\,\ref{fig:JEH}. In Fig.\,\ref{fig:JEH}, we also show the $J$-$E$ characteristics and $\expval{\bar{q}q}$ as a function of $J$. The arrows in the plots indicate the first-order transition point based on the assumption that the stable state has the lowest thermodynamic potential.
As we mention in the main text, when $B$ is small, the flat branch is always favored as shown in Fig.\,\ref{fig:JEH} for $B=14$. However, if we increase $B$, the bending branch gradually becomes stable as shown in Fig.\,\ref{fig:JEH} for $B=16$. This indicates that the first-order phase transition appears at the specific value of $B=B_{\rm gap}$. For larger $B$, the bending branch is always stable compared to the flat branch as shown in Fig.\,\ref{fig:JEH} for $B=18$, $B=20$, and $B=22$. Furthermore, the bending branch clearly becomes single-valued for $B=22$. This indicates that the first-order phase transition is changed to the second-order phase transition at specific point $B=B_{\rm TCP}$.

We have checked that the thermodynamic potential in the broken phase is smaller than or equal to that of the symmetric phase when $B\geq B_{\rm TCP}$.
It should be noted that the critical point in this case is the point where the dotted line intersects with the solid line in the $J$-$\expval{\bar{q}q}$ graphs of Fig.\,\ref{fig:JEH} for $B\geq B_{\rm TCP}$. This is because the two branches for $J<J_{c}$ merge into a single branch for $J\geq J_{c}$ and the thermodynamic potentials of the two branches agree with each other where they merge. This property is robust against a small change of definition of the thermodynamic potential. For example, one may think that another natural choice of the integration range in Eq.\,(4) in the main text may be $0<z<z_{H}$ instead of $0<z<z_{*}$. If we employ this new integration range, the thermodynamic potential will be altered. However, we have checked that the difference for our setup is small enough not to change the property mentioned above: the thermodynamic potential in the broken phase is still less than that in the symmetric phase and the two agree with each other at the point they merge.

\end{document}